\def\centreline{\centerline}
\def\nex{\par\noindent\hang}
\raggedbottom


\def\nex{\par\noindent\hang}

\null
\centerline{\bf The Theoretical Mass--Magnitude Relation of Low-Mass Stars}
\medskip
\centreline{\bf and its Metallicity Dependence}

\vskip 10mm
\par
\centerline{\bf Pavel Kroupa$^1$ and Christopher A. Tout$^{2,3,4}$}
\vskip 20mm
\centerline{$^1$Astronomisches Rechen-Institut}
\vskip 2mm
\centerline{M{\"o}nchhofstra{\ss}e~12-14, D-69120~Heidelberg, Germany}
\vskip 2mm
\centerline{e-mail: s48@ix.urz.uni-heidelberg.de}
\vskip 10mm
\centreline {$^2$Space Telescope Science Institute,}
\vskip 2mm
\centreline{3700 San Martin drive, Baltimore, MD~21218, U.S.A.}
\vskip 2mm
\centerline{e-mail: cat@ast.cam.ac.uk}
\vskip 10mm
\centreline{$^3$On leave from University of Cambridge}
\vskip 10mm
\centreline{$^4$Current Address: Konkoly Observatory of the Hungarian
Academy of Sciences,}
\vskip 2mm
\centreline{H-1525 Budapest, P.O.B.~67, Hungary}

\vfill

\centerline{MNRAS, in press}
\vfill\eject

\hang{
\bf Abstract.
\rm
\noindent
We investigate the dependence of theoretically generated
mass--(absolute magnitude) relations on stellar models.  Using
up-to-date physics we compute models of stars in the mass range $0.1<m
\le 1\,m_\odot$.  We compare the solar-metallicity models with our
older models and also with recent models computed by others.  We
further compare with an empirical mass--(absolute magnitude) relation
that best fits the observed data.  At a given mass below
$0.6\,m_\odot$ the effective temperatures differ substantially from
model to model.  However taken individually each set of models is in
good agreement with observations in the mass--luminosity plane.  A
minimum in the derivative $dm/dM_{\rm V}$ at $M_{\rm V}\approx 11.5$,
which is due to H$_2$ formation and establishment of a fully
convective stellar interior, is present in all photometric bands, for
all models but its position changes from model to model.  This minimum
leads to a maximum in the stellar luminosity function for Galactic
disk stars at $M_{\rm V}\approx11.5, M_{\rm bol}\approx9.8$.  Precise
stellar models should locate this maximum in the stellar luminosity
function at the same magnitude as observations.  This is an extra
constraint on low-mass stellar models.  Models which incorporate the
most realistic theoretical atmospheres and the most recent equation of
state and opacities can satisfy this constraint.  These models are
also in best agreement with the most recent luminosity--(effective
temperature) and mass--luminosity data.  Each set of our models of a
given metallicity (in the range $0.2 > \rm [Fe/H] > -2.3$) shows a
maximum in $-dm/dM_{\rm bol}$, which moves to brighter bolometric
magnitudes with decreasing metallicity.  The change in location of the
maximum, as a function of [Fe/H], follows the location of structure in
luminosity functions for stellar populations with different metal
abundances.  This structure seen in all observed stellar populations
can be accounted for by the mass--luminosity relation and does not
require a maximum in the stellar mass function at
$m\approx0.3\,m_\odot$.

\vskip 5mm
\noindent {\it Subject headings:} stars: low-mass, brown dwarfs --
stars: population~I and~II -- stars: luminosity function, mass function 
-- open clusters and associations: general -- globular
clusters: general

\vfill\eject

\bigskip
\bigbreak
\noindent{\bf 1 INTRODUCTION}
\vskip 12pt
\noindent
It is desirable to understand the shape of the luminosity function in
terms of the underlying mass function.  However, observed luminosity
functions depend, not only on the associated mass functions but also
rather delicately on the mass--luminosity relation of the stars. This
is demonstrated well by Elson et al. (1995, figure~6) and has been
stressed before by D'Antona \& Mazzitelli (1983) and later by Kroupa,
Tout \& Gilmore (1990). The unobservable distribution of stellar
masses is mapped to the observable distribution of stellar
luminosities by multiplying the former by the derivative of the
mass--luminosity relation.

D'Antona \& Mazzitelli (1994) point out that significant uncertainties
remain in our understanding of stellar physics for stars less massive
than about $0.6\,m_\odot$ because atmospheric convection and opacities
are not yet fully understood.  Further discussion of the uncertainties
in stellar structure theory that result from different treatments of
the surface boundary condition is given by Chabrier, Baraffe \& Plez
(1996).  Despite these uncertainties, models of low-mass stars with a
fixed chemical composition demonstrate that the absolute magnitude is
not a simple linear function of stellar mass because opacities, the
equation of state and consequently stellar structure change
significantly with mass.  The derivative $dm/dM_{\rm V}$ varies
substantially with absolute visual magnitude $M_{\rm V}$.  It has a
local maximum at $M_{\rm V}\approx7-8$ and a pronounced minimum at
$M_{\rm V}\approx 11.5$, for solar metallicity stars (Kroupa, Tout \&
Gilmore 1990).  These features coincide, respectively, with a plateau
or depression at $M_{\rm V}\approx7$ and a pronounced maximum at
$M_{\rm V}\approx 11.5$ observed in the stellar luminosity function.
D'Antona \& Mazzitelli (1983) noted that the flattening in the
luminosity function at $M_{\rm V}\approx6-9$ is probably due to the
existence of an inflection in the mass--luminosity relation at a mass
near $m=0.6\,m_\odot$.  Copeland, Jensen \& Jorgensen (1970)
constructed stellar models in the mass range $0.25-2.5\,M_\odot$, and
attributed this increased luminosity near $m=0.5-0.7\,M_\odot$ to the
onset of H$_2$ formation in low-mass stars. They showed how the
associated increase in effective temperature $T_{\rm eff}$ leads to a
break in the main sequence in the HR~diagram at log$_{10}(T_{\rm
eff}/K)\approx3.6$.  More detailed modelling led Kroupa, Tout \&
Gilmore (1990) to suggest that this flattening in the luminosity
function at $M_{\rm V}\approx7$ ($m\approx0.7\,m_\odot$) is caused by
increased importance of H$^-$ opacity in low-mass stars as the stellar
mass is reduced, and that the {\it onset} of H$_2$ formation in a
thin outer shell at $m\approx0.5\,m_\odot$ (their figs.~2 and~3) leads
to an increase in $-dm/dM_{\rm V}$ with increasing $M_{\rm V}$ until a
maximum at $M_{\rm V}\approx11.5$ ($m\approx 0.35\,m_\odot$) is
reached, where the stars are also fully convective.  They derived an
empirical shape for the $m(M_{\rm V})$ relation by comparing the
observed luminosity function with smooth mass functions.  Recent
mass-luminosity data (Henry \& McCarthy 1993) is in good agreement
with this shape.

If the original assertion by Kroupa, Tout \& Gilmore (1990), that the
peak in the luminosity function at $M_{\rm V}\approx11.5$ is due to
the minimum in $dm/dM_{\rm V}$, is true then similar structure should
exist in the luminosity function of different stellar populations , as
pointed out by Kroupa, Tout \& Gilmore (1993).  Further, Kroupa
(1995a) stresses that the amplitude and location of the peak in the
stellar luminosity function may be used as an additional constraint on
stellar models.  Indeed observations with the Hubble Space Telescope
show that the stellar luminosity functions in the three globular
clusters NGC~3697, M15 and 47 Tucanae (Paresce, De Marchi \&
Romaniello 1995, De Marchi \& Paresce 1995a, 1995b, respectively) also
have maxima.  These lie at luminosities somewhat brighter than the
luminosity of the maximum in the solar neighbourhood luminosity
function.

Here we investigate both the extra constraint imposed on population~I
models by the shape of the luminosity function and the sensitivity of
the mass--(absolute magnitude) relation to changes in metallicity.  Of
interest is the slope of the $m(M_{P})$ relation, where $P$ represents
an arbitrary magnitude band, near $m=0.35\,m_\odot$ so we need a fine
grid of stellar models in the mass range $0.1-0.5\,m_\odot$.  We use
the latest version of the Eggleton evolution code (Pols et al. 1995;
Tout et al. 1996) to compute a new set of low-mass stellar models. The
code incorporates substantial recent improvements in our understanding
of the physics of stellar interiors, particularly the opacity and the
equation of state.  We then compare with our computations based on
older opacities and equation of state and with models recently
computed by others (Baraffe et al. 1995; D'Antona \& Mazzitelli 1994).

In Section~2 we introduce the stellar models.  They are compared with
the observations of Galactic disk stars in Section~3. The implications
for the population-II stellar luminosity function are considered in
Section~4.  In Section~5 we summarise our conclusions.

\bigskip
\bigbreak
\noindent{\bf 2 STELLAR MODELS}
\vskip 12pt
\noindent
In addition to our own models we consider the work of D'Antona \&
Mazzitelli (1994) and Baraffe et al. (1995).  We need stellar models
with masses in the range $0.1\,m_\odot \le m\le 0.5\,m_\odot$ so that
we can quantify the first derivative of the mass--(absolute magnitude)
relation near $m = 0.35m_\odot$.  Although other, commonly used models
are to be found in the literature they are not sufficiently well
defined for these purposes.

\vskip 10pt\bigbreak
\noindent{\bf 2.1 Our New and Old Models}
\nobreak\vskip 10pt\nobreak
\noindent
We make our stellar models with the Eggleton evolution code (Eggleton
1971, 1972, 1973).  The most recent version of this code is described
by Pols et al. (1995).  It incorporates new nuclear reaction rates
(Caughlan \& Fowler 1988) and opacities (Iglesias, Rogers \& Wilson
1992, OPAL; Alexander \& Ferguson 1994).  It also uses a new equation
of state (described in detail by Pols et al.)  which treats ionization
equilibria, including pressure ionization, somewhat more precisely and
includes Coulomb interactions.  This equation of state fits well with
that used by Iglesias, Rogers and Wilson for their opacity
computations.  The code has been further improved by incorporation of
bicubic spline interpolations in opacity (Tout et al. 1996).  Our
population~I models have a metallicity $Z = 0.02$, a hydrogen
abundance $X = 0.7$ and a helium abundance $Y = 0.28$.  We use
$\alpha=2$ for the mixing length parameter.  In the old models the
zero-age mixture of metals was determined by the tables of Cox \&
Stewart (1970) while for the new models we use the Solar system
meteoritic mixture determined by Anders \& Grevesse (1989). Accurate
analytic fits to the new models are given by Tout et al. (1996).

Below we show that these new models do not fit the population~I
observations as well as the old models computed by Kroupa, Tout \& 
Gilmore (1990) that used Cox \& Stewart (1970) opacities, reaction
rates from Caughlan, Fowler \& Zimmerman (1975) and an equation of
state based on that of Eggleton, Faulkner \& Flannery (1973), with
pressure ionization described by Hjellming (1989) and a simple
treatment of hydrogen molecules (Webbink 1975).  Both the old and the
new models use a simple Eddington approximation (Woolley \& Stibbs
1953), rather than a stellar atmosphere calculation, for the surface
boundary conditions.

Keeping this in mind, we use the new and old models to illustrate the
differences in the HR diagram, in the $m(M_{P})$ and the
$dm/dM_{P}(M_{P})$ planes, that result from changes in the input
physics. This allows us some insight into which observational
constraint tests which aspect of the input physics.  We follow the
less established convention of writing mass as a function of magnitude
because we are trying to identify the mass of a star with a particular
observed magnitude.  Also, by computing our own models, we can define
a high resolution on the mass grid, which is necessary when studying
the derivative of the mass--luminosity relation. We compute 300
stellar models with masses $0.1\le m\le 1\,m_\odot$ distributed
uniformly in log$_{10}m$.

\bigbreak
\vskip 10pt
\noindent{\bf 2.2 Models by D'Antona \& Mazzitelli}
\nobreak\vskip 10pt\nobreak
\noindent
The greatest deficiency in our models lies in our treatment of the
surface boundary conditions.  This is partially remedied in the work
of D'Antona \& Mazzitelli (1994, also private communication) who use a
grey-body atmosphere integration to an optical depth of $\tau =
2/3$. They compute four sets of models using mixing length theory
(MLT) or their alternative treatment of convection, and the Alexander
(Alexander \& Ferguson 1994) or Kurucz (1991) opacities.  We consider
only the models that have Alexander opacities at low temperatures and
MLT, because (i) the Kurucz opacities do not include molecules and
(ii) the alternative treatment of convection together with Alexander
opacities gives main-sequence stellar models that are virtually
indistinguishable from models computed using MLT and Alexander
opacities.  To ensure arrival on the main sequence we use their
$10^9\,$yr isochrone.  The models listed in their tables~3 and~7 are
spaced by $0.1\,m_\odot$ for $m\ge0.2\,m_\odot$, so that the
resolution of the slope of the mass--(bolometric magnitude) relation,
$dm/dM_{\rm bol}$, in the critical region around $0.35\,m_\odot$ is
limited.

\bigbreak
\vskip 10pt
\noindent{\bf 2.3 Models by Baraffe et al.}
\nobreak\vskip 10pt\nobreak
\noindent
Baraffe et al. (1995) substantially improve the atmospheric modelling.
They allow for convection to extend into the atmosphere and find that
it can indeed reach to an optical depth $\tau \ll 1$.  Chabrier et
al. (1996) discuss the effects of different atmospheric models on
stellar modelling.  Baraffe et al. tabulate their models in their
table~1 with a resolution of $0.1\,m_\odot$ in mass, which again
limits definition of the maximum in $-dm/dM_{\rm bol}$.

\bigbreak
\bigskip
\bigbreak
\noindent{\bf 3 POPULATION I STARS: COMPARISON WITH OBSERVATIONAL
CONSTRAINTS}
\nobreak
\vskip 12pt
\nobreak
\noindent
We can now compare the stellar models introduced in Section~2 with
observations in the hope of learning which physics (opacities,
equation of state, treatment of the atmosphere) leads to stellar
models that are most consistent with available observational
constraints.

\bigbreak\vskip 10pt\bigbreak
\noindent{\bf 3.1 The mass--(absolute magnitude) relation}
\nobreak\vskip 10pt\nobreak
\noindent
Taken alone each set of models is in rather good agreement with
observations in the mass--(bolometric magnitude) plane. In Fig.~1 we
plot the $m(M_{\rm bol})$ data for each series (with $M_{\rm
bol}=4.72$ for the Sun). At a given $M_{\rm bol}<8$, the different
models have masses differing by less than 5\%. For $M_{\rm bol}>11$
the difference in masses amounts to less than $0.02\,m_\odot$.  In the
intermediate region ($8 < M_{\rm bol} < 11$) the models disagree more
substantially.  For completeness we mention that Burrows et al. (1993)
publish models for very low-mass stars ($m \le 0.2\,m_\odot$) that are
indistinguishable from the Baraffe et al. (1995) models in Fig.~1. We
refer the reader to Leggett et al. (1996) for a comparison of the
models published by Burrows et al. with other models in the HR
diagram.

Below a mass of about $0.4\,m_\odot$ stellar structure changes
significantly because stars become fully convective and hydrogen
associates to H$_2$ in the outer layers (Copeland et al. 1970; Kroupa,
Tout \& Gilmore 1990; Baraffe \& Chabrier 1996). Increasing the
stellar mass we find that near $0.35\,m_\odot$ an increase in energy
production leads to dissociation of surface H$_2$ rather than to an
increase in luminosity, and the equilibrium structure adjusts
accordingly.  Suppression of H$_2$ formation (see Kroupa, Tout \&
Gilmore 1990) leads to full convection at a significantly lower
stellar mass ($m\approx 0.25\,m_\odot$).  This illustrates the
importance of H$_2$ formation for convection.  The region in mass
around $0.35\,m_\odot$ is critical and additional observational
mass--luminosity data is indispensable for constraining the theory of
low-mass stellar structure. The derivative $dm/dM_{\rm bol}$
emphasizes the differences between models and is a potentially
powerful diagnostic.

In Fig.~2 we compare mass--luminosity observations with our new and
old models and with the empirical $m(M_{\rm V})$ relation derived by
Kroupa, Tout \& Gilmore (1993). Theoretical $M_{\rm V}$ values are
estimated from $M_{\rm bol}$ following Kroupa, Tout \& Gilmore
(1990). The agreement of the models with the data can be estimated by
computing the chi-squared statistic for $M_{\rm V}>7.0$ (26 data
points): $\chi^2=186$ for the old models, $\chi^2=132$ for the Kroupa,
Tout \& Gilmore (1993) relation and $\chi^2=57$ for the new models.
Significant improvement in the $m(M_{\rm V})$ plane is obtained with
the new physics.  However, in making this comparison, we must be aware
that the model $m(M_{\rm V})$ relations are valid for a single-age,
single-metallicity stellar population and that they have been computed
with the rather uncertain bolometric corrections.  The observational
data, on the other hand, represent a mixture of Galactic disk stars of
different metal abundance and age.

\bigbreak
\vskip 10pt
\noindent{\bf 3.2 The first derivative of the mass--(absolute magnitude)
relation}
\nobreak\vskip 10pt\nobreak
\noindent
Returning to the critical region around $m=0.35\,m_\odot$, we evaluate
$dm/dM_{P}$ for the old and new stellar models and in different
photometric bands for the Kroupa, Tout \& Gilmore (1993) relation.
Absolute magnitudes in the photometric J-, H- and K-bands are obtained
for each mass from the colour--magnitude relations given by
equation~(1) of Henry \& McCarthy (1993). To obtain absolute I-band
magnitudes we select a subset of $M_{\rm V}$, $V-I$ data from fig.~10
of Monet et al. (1992) which satisfy $M_{\rm V} > 3.96 +
3.34\,(V-I)$. This eliminates obvious subdwarfs.  A linear regression
on the data, with each point weighted by its stated error, gives

$$ V-I = 0.308\,M_{\rm V} - 1.046, \eqno (1)$$

\noindent which is very similar to the relation derived by Stobie,
Ishida \& Peacock (1989)

$$V-I = 0.299\,M_{\rm V} - 0.865.  \eqno (2)$$

\noindent We also derive a cubic weighted least squares relation

$$V-I = 2.696\times10^{-3}\,M_{\rm V}^3 - 9.857\times10^{-2}\,M_{\rm
V}^2 + 1.433\,M_{\rm V} - 4.949. \eqno (3)$$

\noindent 
These three $V-I, M_{\rm V}$ relations allow us to estimate the
uncertainties associated with the derivative of the colour-magnitude
relation which also enters $dm/dM_{P}$. In particular, equation~3
allows for the steepening of the $M_{\rm V}(V-I)$ relation near
$V-I=2.8$ evident in Fig.~3, in which we plot the colour--magnitude
data together with the three relations. This steepening results from
the changes in stellar structure discussed in Section~1 and has also
been discussed by Leggett, Harris \& Dahn (1994).

The derivatives of the mass--(absolute magnitude) relations as
functions of $M_{\rm V}$ are shown in Fig.~4.  We also plot
$dm/dM_{P}$ for the Kroupa, Tout \& Gilmore (1993) $m(M_{\rm V})$
relation in the I-, J-, H- and K-bands. The data are tabulated in the
appendix (with equation~1 to transform from $M_{\rm V}$ to $M_{\rm
I}$).  Conversion to $M_{\rm bol}$ from $M_{\rm V}$ for our empirical
relation is made with $V-I$ from equation~1 and bolometric corrections
from equation~6 of Monet et al. (1992).

The minimum in the derivative of the Kroupa, Tout \& Gilmore (1993)
$m(M_{\rm V})$ relation is pronounced in all photometric bands and
agrees in location and amplitude with the observed photometric stellar
luminosity functions in the V- and I-bands. This is no surprise in the
V-band because it was derived to maximize the agreement with the
observational $m(M_{\rm V})$ data provided by Popper (1980), and that
the amplitude and position of the maximum in $-dm/dM_{\rm V}$ fit the
stellar luminosity function.  The luminosity function used here is the
estimated parent distribution of Malmquist corrected photometric
luminosity functions derived by Kroupa (1995b).  This is based on
estimates of stellar space densities by photometric parallax from
low-spatial resolution but deep pencil-beam surveys.  It measures the
spatial densities of stellar systems, approximately half of which are
binary stars, and thus underestimates the number of faint stars (Buser
\& Kaeser 1985, Kroupa, Tout \& Gilmore 1991, 1993; Piskunov \& Malkov
1991; Kroupa 1995a).  However the position of the peak in the
luminosity function does not change significantly as a result of this
bias (fig.~20 in Kroupa, Tout \& Gilmore 1993; fig.~1 in Kroupa
1995a). We therefore use the photometric luminosity function, which is
based on 448~stars and is statistically much more well defined than
the nearby luminosity function, to constrain the {\it position} and
approximate {\it amplitude} of the minimum in the first derivative of
the mass--magnitude relation. We cannot completely constrain the shape
of the minimum because this would require detailed knowledge of the
stellar mass function.  By using these average colour--magnitude
relations together with the empirical $m(M_{\rm V})$ relation we are
in effect modelling a single metallicity and single age stellar
population. This is logically consistent with the adoption of the
Malmquist corrected data for the photometric luminosity function
because the stellar samples used to define the relationships and the
luminosity functions all belong to the solar neighbourhood.

Returning to Fig.~4, we see that the stellar models imply a minimum in
$dm/dM_{P}$ which approximately agrees in location and amplitude
with the maximum in the observed luminosity functions. This is
evidence that the peak in the stellar luminosity function is caused by
the minimum in $dm/dM_{P}$ rather than a maximum in the stellar
mass function. However the new stellar models place the peak at an
absolute V-band magnitude which is too bright by about $0.8\,$mag,
while the old models were in better agreement with the data.

We first consider changing the two free parameters of stellar
evolution: hydrogen abundance $X$ (or equivalently helium, $Y$) and
the ratio $\alpha$ of the mixing length to the pressure scale height.
So as not to be hindered by uncertainties in the $M_{\rm bol}-M_{\rm
V}$ relation, we compare the models in bolometric magnitudes and take
the best estimate for the photometric luminosity function in
bolometric magnitudes obtained by Kroupa (1995b) as the observational
data.  We use $M_{\rm bol}=-2.5\,{\rm log}_{10}(L/L_\odot) + 4.72$,
where $L$ is the stellar luminosity.  From Fig.~5 we deduce that
varying $\alpha$ with fixed $X$ has a negligible effect, while varying
$X$ with fixed $\alpha$ increases the maximum in $-dm/dM_{\rm bol}$
for models with more hydrogen and moves it to slightly brighter
magnitudes.  Reasonable variation in $\alpha$ and $X$ cannot bring the
maximum in $-dm/dM_{P}$ for our (improved) models into agreement with
the observations.  Thus, although the new models have an improved
chi-squared fit w.r.t. the new mass--(absolute magnitude) data, the
location of the minimum in the derivative is too bright by about~0.8
mag in the V-band and about $0.5\,$mag in $M_{\rm bol}$.

To understand why this is, we compare our stellar models with those of
Baraffe et al. (1995) and D'Antona \& Mazzitelli (1994).  We evaluate
$M_{\rm bol}$ and fit cubic splines to their $10^9\,$yr isochrones.
Derivatives $dm/dM_{\rm bol}$ are calculated from the fitted
cubic-spline relations and we plot them in Fig.~6.  We make the
following observations: {\bf (a)} our new models and those of D'Antona
\& Mazzitelli (1994) place the minimum in $dm/dM_{\rm bol}$ at about
the same $M_{\rm bol}\approx9$; {\bf (b)} the peak in our models is
more pronounced; we attribute this to our finer resolution in the mass
grid; {\bf (c)} our old models have the minimum at $M_{\rm
bol}\approx9.5$, in agreement with the observations, as already noted
above; and {\bf (d)} the most modern stellar models computed by
Baraffe et al. (1995) better reproduce the location of the maximum in
the luminosity function; the amplitude we derive for their models is
again limited by the low resolution of their mass grid.

We find it unlikely that structure would exist in the mass function so
close to the point where similar structure would result directly from
the mass--luminosity relation.  The results of this section and
section~4 lend credence to the simplest hypothesis that the peak in
the luminosity function results from the maximum in $-dm/dM_{P}$.
This hypothesis was basic to the investigation reported in Kroupa,
Tout \& Gilmore (1990) and used in subsequent work.  A mass function
with a maximum is also inconsistent with star-count data (Section~9 in
Kroupa, Tout \& Gilmore 1993).  Thus we conclude that fortuitous
cancellation of errors in stellar structure led to good agreement
between our old models and data.  It is only when both interior and
atmospheric physics are improved (Baraffe et al. 1995) that we can
improve on this fit.

\vskip 10pt\bigbreak
\noindent{\bf 3.3 The Hertzsprung-Russel diagram}
\nobreak\vskip 10pt\nobreak
\noindent
We have compared stellar models with observations in the
mass--luminosity plane and invoked the first derivative of the
mass--(absolute magnitude) relation as an additional constraint.  We
now extend the comparison to the luminosity--(effective temperature)
plane.  Fig.~7 is the HR diagram for low-mass stars.  We take our data
from Popper (1980) and Kirkpatrick et al. (1993).  The latter describe
how their estimates of the effective temperature are hotter than
previous empirical estimates by about $300\,$K for $M_{\rm
bol}\approx9.7$ ($m\approx0.3\,m_\odot$) and by about $350\,$K at
$M_{\rm bol}\approx12.2$ ($m\approx0.1\,m_\odot$).  Unfortunately no
independent (i.e. from binary star orbits) stellar mass estimates are
available for the data published by Kirkpatrick et al. (1993) and we
believe they are now superseded by a new empirical determination of
$L(T_{\rm eff})$ data by Leggett et al. (1996). These data, for
Galactic-disc stars, are in good agreement with those of Popper
(1980), lying only about 100~K cooler for $-2.6<$
log$_{10}(L/L_\odot)<-1.9$.

All stellar models suggest that the effective temperature increases
more slowly with increasing luminosity in the range $3.46 <
\log_{10}(T_{\rm eff}/{\rm K}) < 3.60$.  This is due to the
dissociation of H$_2$ as $T_{\rm eff}$ increases and the related
establishment of a radiative stellar core in stars more massive than
about $0.35\,m_\odot$ and is associated with the minimum in $dm/dM_P$.
Models of D'Antona \& Mazzitelli (1994) are in agreement with the
Kirkpatrick et al. (1993) data.  Our new models also agree with this data
down to about $0.2\,m_\odot$ but are too cool at lower
masses. However, the models computed by Baraffe et al. (1995) using
the most realistic treatment of the atmosphere are too cool by
log$_{10}T_{\rm eff}\approx 0.04$ with respect to the Kirkpatrick et
al. data, but are in agreement with the older data published by Popper
(1980) and the most recent compilation by Leggett et al. (1996).  In
the $M_{\rm V}, V-I$ diagram (figs~1 and~2 in Baraffe et al. 1995)
their models are bluer (i.e. hotter) than observational data.  Baraffe
et al. (1995) discuss the reasons for this.

\bigbreak
\vskip 10pt
\noindent{\bf 3.4 Summary}
\nobreak\vskip 10pt\nobreak
\noindent
We summarise our comparison of stellar models with
observed data in Table~1.

\bigbreak
\vskip 3mm
\bigbreak

\noindent {\bf Table 1:} Comparison with observational data

\nobreak
\vskip 1mm
\nobreak
{\hsize 14 cm \settabs 10 \columns

\+model &&$m(M_{\rm V})$  &&HR diagram  &&&&&$-dm/dM_{\rm bol}$ \cr

\+\cr

\+our old   &&OK &&too hot for           &&&&&OK\cr
\+          &&   &&log$_{10}T_{\rm eff}<3.52$ \cr
\+\cr
\+our new   &&OK &&too hot for &&&&&too bright \cr
\+          &&   &&$3.52<$log$_{10}T_{\rm eff}<3.58$ \cr
\+\cr
\+D'Antona  &&OK &&too hot for          &&&&&too bright \cr
\+          &&   &&log$_{10}T_{\rm eff}<3.58$ \cr
\+\cr
\+Baraffe   &&OK &&OK                   &&&&&OK \cr
  
} 
  
\bigbreak\vskip 3mm

\noindent 
Column~1 lists the models (see Section~2) and column~2 lists our
assessment of their fit to the $m(M_{\rm V})$ data plotted in Fig.~2
(see also Fig.~1). Column~3 is our assessment of the fit to the
luminosity--(effective temperature) data of Leggett et al. (1996) and
Popper (1980).  Consistency with the locations of the maximum in
$-dm/dM_{\rm bol}$ and of the peak in the stellar luminosity function
from Kroupa (1995b) is described in column~4.

We conclude that the most sophisticated models by Baraffe et
al. (1995) provide the overall best agreement with all recent
observational constraints. The same conclusion concerning the HR
diagram is also arrived at by Leggett et al. (1996).  {\it All} models
provide reasonable fits to the $m(M_{\rm V})$ data which, without
further mass determinations around $0.35\,m_\odot$, is not sufficient
to distinguish between models.

\bigbreak
\bigskip
\noindent{\bf 4 IMPLICATIONS FOR LUMINOSITY FUNCTIONS OF POPULATION 
II STARS}
\vskip 12pt
\noindent
Until recently there were essentially no observational constraints on
metal-poor stellar models. Comparison with colour--magnitude data is
hindered by our ignorance of the state of the stellar atmosphere. An
example is to be found in figs~1 and~2 of Baraffe et al. (1995), where
we see that for $M_{\rm V}>15$ the population~I models are too blue by
up to $\Delta (V-I) \approx 0.4$, or conversely, are too bright by
$\Delta M_{\rm V}\approx 1\,$mag for $V-I>3.4$. The population~II
models, however, appear to fit the subdwarf sequence.

We have demonstrated (Section~3.2) that the location of the minimum in
$dm/dM_{P}$ can be used as a constraint on stellar models.  We can
also apply this test to population~II stars by comparing the position
of the theoretical minimum in the derivative of the mass--magnitude
relation with the magnitude of the turnover in observed stellar
luminosity functions.  Testing metal-poor luminosity functions is also
an important step towards establishing whether the mass function
varies with star-formation conditions.  The behaviour of the location
of the turnover in the luminosity function with metallicity is
conveniently summarised and extended to Galactic clusters with solar
abundance by von Hippel et al. (1996).  In their fig.~5 they show how
the position of the turnover moves to brighter magnitudes for more
metal deficient stellar populations.  They compare this with the
constant mass locus of stellar models and find acceptable
agreement. This suggests that the change in location of the structure
in the stellar luminosity function with metallicity may be due to
changes in stellar physics rather than the stellar mass function. By
considering constant mass loci alone we cannot however establish that
the location of the minimum in the derivative of the mass--magnitude
relation also behaves like the observational sample.  In order to test
this in a more consistent way we compare empirical data with the
theoretical position of the minimum in $dm/dM_{\rm bol}$.

\bigbreak
\vskip 10pt
\noindent{\bf 4.1 Theoretical data}
\nobreak\vskip 10pt\nobreak
\noindent
For the theoretical data set we compute new stellar models for
metallicities $Z=0.03$, 0.02, 0.01, 0.004, 0.001, 0.0003 and~0.0001
with hydrogen abundance $X=0.76 - 3Z$ and helium abundance $Y=0.24 +
2Z$, and plot some of the first derivatives $dm/dM_{\rm bol}$ in
Fig.~8.  All have maxima in $-dm/dM_{\rm bol}$ which are at a brighter
magnitude $M_{\rm bol}^{\rm max}$ for lower metal abundance.  This
effect is less significant near the solar value.  For a given mass,
metal-poorer stars are brighter and a little smaller.  Consequently
hydrogen dissociation in the outer layers takes place at slightly but
not significantly lower masses.  These low metallicity stars are still
brighter than their metal rich counterparts, and so the inflexion in
the mass--magnitude relation moves to higher luminosities with
decreasing metallicity.  We plot the resulting $M_{\rm bol}^{\rm
max}({\rm [Fe/H]})$ relationship in Fig.~9 with filled circles.

Recall that for solar metallicities our models predict an $M_{\rm
bol}^{\rm max}$ that is too bright by about 0.8~mag
(Section~3.2). However, as indicated by our modelling shown in Fig.~5,
we expect the change in $M_{\rm bol}^{\rm max}$ with [Fe/H] to be
more robust to changes in stellar physics than the actual position of
the maximum $M_{\rm bol} ^{\rm max}$, because it depends primarily on
$Z$.  Available theoretical work (see tables~1--4 in D'Antona
\& Mazzitelli 1996) suggests that the bolometric corrections for the
I-band do not drastically change with metal abundance so that we do
not transform our $M_{\rm bol}^{\rm max}$ values to the I-band.  We
see that this approximation is sufficient for the present purpose by
comparing with theoretical $M_{\rm I}^{\rm max}({\rm [Fe/H]})$ values
from D'Antona \& Mazzitelli (1996).  When we compute our own models we
can define an arbitrarily high resolution in the mass grid and
metallicity grid.  However we are limited by not being able to
incorporate a realistic stellar atmosphere.  D'Antona \& Mazzitelli
(1996) have also computed stellar models for different metallicities
and from their fig.~7 we obtain $M_{\rm I}^{\rm max}({\rm [Fe/H]})$.
They plot theoretical I-band luminosity functions for a constant mass
function and from these we compute $-dm/dM_{\rm I}$. Given the sparse
grid in stellar masses and the wide magnitude bins we estimate the
likely range for the maximum in $-dm/dM_{\rm I}$ from the magnitude
bin width and thus obtain lower and upper limits on $M_{\rm I}^{\rm
max}({\rm [Fe/H]})$. These are plotted in Fig.~9 as two sets of open
squares.  Both limiting curves have been shifted by $+0.7$~mag because
their models also place the minimum at too bright a magnitude
(Fig.~6).

\bigbreak
\vskip 10pt
\noindent{\bf 4.2 Observational data}
\nobreak\vskip 10pt\nobreak
\noindent
Good observations of four globular clusters are now available. The
stellar luminosity functions extend to the vicinity of the
hydrogen-burning mass limit and are defined by a few thousand stars.
Three of these luminosity functions show the expected structure: the
luminosity function of NGC~6397 ([Fe/H] $=-1.91$) has a sharp peak at
$M_{\rm I}^{\rm max}=8.55$ (Paresce et al. 1995); for M15 ([Fe/H]
$=-2.26$) it has a somewhat broader peak at $M_{\rm I}^{\rm max}=8.60$
(De Marchi \& Paresce 1995a); and for 47~Tuc ([Fe/H] $=-0.65$) it has
a pronounced peak at $M_{\rm I}^{\rm max}=8.90$ (De Marchi \& Paresce
1995b).  However the luminosity function of the dynamically young
cluster $\omega$~Cen ([Fe/H] $=-1.60$, with a substantial spread of
abundances), on the other hand, rises to $M_{\rm I}=8.61$ and remains
flat at lower luminosities (Elson et al. 1995).  Because this
luminosity function does not show a pronounced maximum it merits
future investigation.  

To these globular clusters we add the Galactic clusters NGC~2420,
NGC~2477, Pleiades and Hyades, and the solar neighbourhood sample.
The luminosity function of NGC~2477 flattens without a peak in a
similar way to that of $\omega$~Cen.  For the solar neighbourhood we
use the parent distribution of Malmquist corrected photometric
luminosity functions estimated by Kroupa (1995b) from 448 stars within
a distance of $100 - 200\,$pc of the Sun.  We assign it an average
${\rm [Fe/H]}=-0.2$ (Edvardsson et al. 1993, Wyse \& Gilmore 1995).
We have already plotted this luminosity function in the V-band
magnitudes (Fig.~4) and in bolometric magnitudes (Fig.~5).

We illustrate the uncertainties in determining $M_{\rm I}^{\rm max}$
owing to magnitude binning by marking the bin-boundary magnitudes that
bracket the position of the turnover in Fig.~9.  For the solar
neighbourhood luminosity function we estimate the likely range of
$M_{\rm I}^{\rm max}$ from the bin boundaries in the V-band (Kroupa
1995b) and equation~1 and note that reasonable changes in the $M_{\rm
V}(V-I)$ relation do not significantly affect the estimated values
(Fig.~4).  For the two luminosity functions that do not show a
pronounced maximum we mark the bright bin boundary before the
flattening and indicate the next faint bin boundary by an arrowhead.

We divide the sample into two groups: those which are defined by more
than a few hundred stars are shown as thick open circles; those with
less are shown as thin open circles. These are NGC~2477 (based on 53
stars) and NGC~2420 (only 19 stars) from von Hippel et al. (1996). 

The Pleiades ([Fe/H]$=+0.03$) luminosity function (Hambly, Hawkins \&
Jameson 1991) has a well defined shape with a maximum, but it is not
clear how old the low-mass stars are. The canonical age is about
$7\times10^7$~yr, but by studying Li depletion Basri, Marcy \& Graham
(1996) suggest that it may be $1.2\times10^8$~yr. Using the pre-main
sequence evolution tracks in table~7 of D'Antona \& Mazzitelli (1994)
the former age implies that a star with a mass of $0.3\,m_\odot$ lies
above the main sequence by $\delta M_{\rm bol}\approx0.37$~mag and the
latter estimate implies $\delta M_{\rm bol}\approx0.14$~mag.  The
distance modulus is usually taken to be~5.5~mag but Gatewood et
al. (1990) measure~$5.9\pm0.26$~mag.  Further discussion of the
possible uncertainties is given in section~6.2.1 of Kroupa (1995c).
In Fig.~9 we portray these uncertainties by plotting $M_{\rm I}^{\rm
max}$ as thin open circles for the various combinations of age and
distance modulus.  

The Hyades data point is taken from the estimate by Reid (1993) of the
luminosity function based on stars with a membership probability of at
least 50~per cent (panel~a of his fig.8). For the transformation from
the V-band to the I-band we use the $M_{\rm V}(V-I)$ relationship
derived by Reid for the Hyades stars because the cluster is relatively
metal rich ([Fe/H]$=+0.15$).

Dynamical evolution of clusters may affect the peak magnitude because
low-mass stars preferentially evaporate as a cluster evolves.  However
the predicted value is quite insensitive to reasonable changes in a
power-law mass function (fig.~4 of Kroupa \& Gilmore 1994).  Figs.~10
and~12 of Kroupa (1995c) show that as a cluster looses low-mass stars
to near exhaustion, the peak in the stellar luminosity function moves
to a brighter magnitude by only about $0.5\,$mag in $M_{\rm V}$ or
$0.29\,$mag in $M_{\rm bol}$.  Because this is smaller than the
difference between the high- and low-metallicity clusters plotted in
Fig.~9, and because we do not expect any of these clusters to have
lost all of their low-mass stars, dynamical evolution does not affect
our conclusions.

\bigbreak
\vskip 10pt
\noindent{\bf 4.3 Interpretation}
\nobreak\vskip 10pt\nobreak
\noindent
While the observational sample has large uncertainties, the difference
between the high metallicity luminosity functions (Galactic-disc
photometric luminosity function, Hyades, Pleiades, NGC~2420, NGC~2477)
and the three low-metallicity clusters (M~15, $\omega$~Cen, NGC~6397)
confirms that there is a correlation of peak magnitude with
metallicity (Fig.~9). Stellar populations with higher metal abundances
tend to have a maximum or turnover in the stellar luminosity function
that lies at fainter absolute magnitudes.

Comparison of our theoretical $dM_{\rm bol}^{\rm max}/d{\rm
[Fe/H]}({\rm [Fe/H]})$ relationship with $dM_{\rm I}^{\rm max}/d{\rm
[Fe/H]}({\rm [Fe/H]})$ from D'Antona \& Mazzitelli (1996) shows that
both are in agreement and also shows that the bolometric corrections
for the I-band do not change significantly with metallicity. We do not
have to shift our relationship to fainter magnitudes because the
bolometric corrections and our offset (Fig.~6) are such that $M_{\rm
bol}^{\rm max}({\rm [Fe/H]})$ coincides roughly with the I-band
observational relationship.  The observed trend is consistent with the
hypothesis that the structure in stellar luminosity functions near
$m=0.35\,m_\odot$ reflects structure in the first derivative of the
mass--absolute-magnitude relation and not in the mass function.

\bigbreak
\noindent{\bf 5 CONCLUSIONS}
\vskip 12pt
\noindent
All the models we have tested are in good individual agreement with
data in the mass--luminosity plane (Figs. 1 and~2).  The largest
differences between models occur near a mass of $0.35\,m_\odot$,
where, as the mass falls, stellar structure changes significantly
owing to the onset of full convection and formation of H$_2$ near the
surface. Additional observational mass--luminosity data are required
around this critical mass in order to better constrain models. These
constraints need not only come from well-determined binary-star orbits
but may also be derived from gravitational microlensing of distant
stars by high-proper motion (i.e. nearby) stars (Paczy\'nski 1995).
The theoretical and empirical HR diagram reflects the changes in
stellar interiors at the critical mass (Fig.~7)

Following Kroupa, Tout \& Gilmore (1990), we argue that the peak in
the stellar luminosity function at $M_{\rm V}\approx11.5$ ($m\approx
0.35\,m_\odot$) for population~I stars results from the minimum in
$dm/dM_{P}$. This gives us an additional constraint on stellar models
because the position of the peak is not significantly dependent on the
precise slope of a smooth mass function.  According to our hypothesis,
precise models should have a maximum in $-dm/dM_{\rm bol}$ at $M_{\rm
bol}\approx 9.8$.  The models computed by Baraffe et al. (1995), who
treat the stellar atmosphere in the most detail, are able to reproduce
the location of this feature in the stellar luminosity function.

In Section~4 we computed population~II stars and found that the
minimum in the first derivative of the mass--magnitude relation moves
to brighter magnitudes with decreasing metallicity.  Our stellar
models for metallicity extending down to [Fe/H] $=-2.3$ reproduce the
behaviour of the position of the peaks and turnovers of the empirical
stellar luminosity functions relative to solar metallicity (Fig.~9).
In this way the position of the peak observed in stellar luminosity
functions for both metal poor and metal rich stars can be accounted
for by the minimum in the first derivative of the mass--(absolute
magnitude) relation (see also von Hippel et al. 1996).  There is no
indication that additional variation of the stellar mass function is
needed, but more conclusive results must await a deeper analysis of
each individual cluster luminosity function.

Despite the recent significant progress in modelling the internal
structure of low-mass stars significant errors remain when a realistic
atmosphere is not included: models with a standard Eddington
approximation for the surface boundary condition (our models) and
those that use a grey atmosphere (D'Antona \& Mazzitelli 1994) are
both in reasonable agreement with mass--luminosity data but place the
maximum in $-dm/dM_{P}$ at too bright a magnitude and are too hot for
log$_{10}(L/L_\odot)<-1.4$.  A realistic atmosphere (Baraffe et
al. 1995) leads to much better agreement with mass--luminosity data,
the estimated maximum in $-dm/dM_{P}$ and the luminosity--(effective
temperature) data for faint stars.

We urge researchers who can compute stellar models with realistic
stellar atmospheres to produce models on a fine grid of masses in the
range $0.1-1.0\,m_\odot$ so as to resolve the extremum in the first
derivative of the mass--magnitude relation near a mass of
$0.35\,m_\odot$.  They should also consider a large range of
metallicities so that their models may be tested against an
observational sample similar to the one presented in Fig.~9.

\bigskip
\bigbreak
\noindent{\bf ACKNOWLEDGEMENTS}
\vskip 12pt
\noindent
We thank F. D'Antona for sending us results of her model calculations,
and Derek Jones for useful suggestions. We also acknowledge the
comments of an anonymous referee.  CAT is very grateful to the UK
PPARC for an Advanced Fellowship during which much of this work was
completed. PK thanks the Institute of Astronomy, Cambridge and the
Space Telescope Science Institute, where most of this work was done,
for hospitality.

\vfill\eject

\bigskip
\noindent{\bf REFERENCES}
\nobreak
\bigskip
\nex Alexander D.R., Ferguson J.W., 1994, ApJ, 437, 879
\nex Anders E., Grevesse N., 1989, Geochim. Cosmochim. Acta, 53, 197 
\nex Andersen J., 1991, A\&AR, 3, 91
\nex Baraffe I., Chabrier G., 1996, ApJ, 461, L51
\nex Baraffe I., Chabrier G., Allard F., Hauschildt P.H., 1995, ApJ, 446, L35
\nex Basri G., Marcy G. W., Graham J. R., 1996, ApJ, 458, 600
\nex Burrows A., Hubbard W.B., Saumon D., Lunine J.I., 1993, ApJ, 406, 158
\nex Buser R., Kaeser U., 1985, A\&A, 145, 1
\nex Caughlan G.R., Fowler W.A., 1988, At. Data Nucl. Data Tables, 40, 284
\nex Caughlan G.R., Fowler W.A., Zimmermann B.A., 1975, ARA\&A, 13, 69
\nex Chabrier G., Baraffe I., Plez B., 1996, ApJ, 459, L91
\nex Copeland H., Jensen J.O., Jorgensen H.E., 1970, A\&A, 5, 12
\nex Cox A.N., Stewart J.N., 1970, ApJS, 19, 243
\nex D'Antona F., Mazzitelli I., 1983, A\&A, 127, 149
\nex D'Antona F., Mazzitelli I., 1994, ApJS, 90, 467
\nex D'Antona F., Mazzitelli I., 1996, ApJ, 456, 329
\nex De Marchi G., Paresce F., 1995a, A\&A, 304, 202
\nex De Marchi G., Paresce F., 1995b, A\&A, 304, 211
\nex Edvardsson B., Andersen J., Gustafsson B., Lambert D. L., Nissen
     P. E., Tomkin J., 1993, A\&A, 275, 101 
\nex Eggleton P.P., 1971, MNRAS, 151, 351
\nex Eggleton P.P., 1972, MNRAS, 156, 361
\nex Eggleton P.P., 1973, MNRAS, 163, 279
\nex Eggleton P.P., Faulkner J., Flannery B.P., 1973, A\&A, 23, 325 
\nex Elson R.A.W., Gilmore G.F., Santiago B.X., Casertano S.,
     1995, AJ, 110, 682
\nex Gatewood G., Castelaz M., Han I., Persinger T., Stein J.,
     Stephensen B., Tangren W., 1990, ApJ, 364, 114
\nex Hambly N.C., Hawkins M.R.S., Jameson R.F., 1991, MNRAS,
     253, 1
\nex Henry T.J., McCarthy D.W., 1993, AJ, 106, 773
\nex von Hippel T., Gilmore G., Tanvir N., Robinson D., Jones D.H.P., 
     1996, AJ, 112, 192
\nex Hjellming, 1989, PhD thesis, University of Illinois
\nex Iglesias C.A., Rogers F.J., Wilson B.G., 1992, ApJ, 397, 717 (OPAL)
\nex Kirkpatrick J.D., Kelly D.M., Rieke G.H., Liebert J., Allard
     F., Wehrse R., 1993, ApJ, 402, 643
\nex Kroupa P., 1995a, ApJ 453, 358
\nex Kroupa P., 1995b, ApJ 453, 350
\nex Kroupa P., 1995c, MNRAS, 277, 1522
\nex Kroupa P., Gilmore G., 1994, MNRAS, 269, 655 
\nex Kroupa P., Tout C.A., Gilmore G., 1990, MNRAS, 244, 76
\nex Kroupa P., Tout C.A., Gilmore G., 1991, MNRAS, 251, 293
\nex Kroupa P., Tout C.A., Gilmore G., 1993, MNRAS, 262, 545
\nex Kurucz R.L., 1991, in Crivellari L., Hubeny I., Hummer D.G., eds,
     NATO ASI Ser., Stellar Atmospheres: Beyond the Classical Models. 
     Kluwer, Dordrecht, p.441
\nex Leggett S. K., Harris H. C., Dahn C. C., 1994, AJ, 108, 944
\nex Leggett S. K., Allard F., Berriman G., Dahn C. C., Hauschildt 
     P. H., 1996, ApJS, 104, 117  
\nex Monet D.G., Dahn C.C., Vrba F.J., Harris H.C., Pier J.R.,
     Luginbuhl C.B., Ables H.D., 1992, AJ, 103, 638
\nex Paczy\'nski B., 1995, Acta Astronomica 45, 345
\nex Paresce F., De Marchi G., Romaniello M., 1995, ApJ, 440, 216
\nex Piskunov A.E., Malkov O.Yu., 1991, A\&A, 247, 87
\nex Pols O.R., Tout C.A., Eggleton P.P., Han Z., 1995, MNRAS,
     274, 964
\nex Popper D.M., 1980, ARA\&A, 18, 115
\nex Reid N., 1993, MNRAS, 265, 785
\nex Stobie R.S., Ishida K., Peacock J.A., 1989, MNRAS, 238, 709
\nex Tout C.A., Pols O.R., Eggleton P.P., Han Z., 1996, MNRAS, 281,257
\nex Webbink R.F., 1975, PhD thesis, University of Cambridge
\nex Woolley R.v.d.R., Stibbs D.W.N., 1953. The Outer Layers of a Star. 
     Clarendon Press, Oxford 
\nex Wyse R. F. G., Gilmore G., 1995, AJ, 110, 2771

\vfill\eject


\vfill\eject

\centerline{\bf Figure captions}
\smallskip

\noindent {\bf Figure 1.}
Mass--(absolute bolometric magnitude) relations for solar-abundance
stellar models. The solid curve represents our new stellar models and
the dashed curve our old models. The models of D'Antona \& Mazzitelli
(1994) are plotted as the open circles and those of Baraffe et
al. (1995) as filled squares

\vskip 5mm

\noindent {\bf Figure 2.}
Comparison of our new (long-dashed curve) and old (short-dashed curve)
solar-abundance stellar models with observations with estimated errors
(Henry \& McCarthy 1993: open circles; Andersen 1991: open squares).
The solid curve is the empirical $m(M_{\rm V})$ relation derived by
Kroupa, Tout \& Gilmore (1993).

\vskip 5mm

\noindent {\bf Figure 3.}
Colour--magnitude data from fig.~10 in Monet et al. (1992).  Obvious
subdwarfs are plotted with asterisks.  Dwarfs are shown as errorbars
(see Section~3.2 for details).  The solid line is our linear
regression (equation~1), the long-dashed line is equation~(2) (from
Stobie et al. 1989) and the short-dashed curve is our cubic fit
(equation~3). The dotted line, $M_{\rm V}=3.96+3.34\,(V-I)$, separates
subdwarfs from dwarfs.

\vskip 5mm

\noindent {\bf Figure 4.}
Derivatives of $m(M_{P})$ relations.  

\noindent (a) 
The visual band $P = \rm V$.  The solid curve is the derivative of
the Kroupa, Tout \& Gilmore (1993) empirical relation.  It peaks at
$M_{\rm V} = 11.6$.  The long-dashed curve is the derivative of our
new models and the short-dashed curve that of our old models.  Filled
circles are derived from observed luminosity functions (Kroupa 1995b)
and scaled to this figure.

\vskip 2mm

\noindent (b)
The I-band.  The three curves are all derivatives of the empirical
Kroupa, Tout \& Gilmore (1993) relation transformed from the V-band:
for the solid curve we used our linear fit (equation~1); for the
dotted curve we used Stobie et al. (1889) linear fit (equation~2); and
for the dashed curve we used our cubic fit (equation~3).  The peak is
at $M_{\rm I} = 9.00$ irrespective of which relation is used.  Open
squares are scaled from the luminosity function of the Pleiades
(Hambly et al. 1991) with a distance modulus of 6.

\vskip 2mm

\noindent (c)
The derivatives of the empirical relation in the K (solid line), J
(dashed line) and H (dotted line) bands.  The peaks are at $M_{\rm
K}^{\rm max} = 6.95$, $M_{\rm J}^{\rm max} = 7.75$ and $M_{\rm H}^{\rm
max} = 7.20$.

\vskip 5mm

\noindent {\bf Figure 5.}
The effects of the mixing length ratio $\alpha$ and hydrogen abundance
$X$ on $dm/dM_{\rm bol}$. The solid line is standard with $\alpha=2$
and $X=0.7$. The dashed curve has $\alpha=1$ ($X = 0.7$) and the
dotted curve has $X=0.8$ ($\alpha = 2$).  The histogram is the best
estimate photometric luminosity function from Kroupa (1995b) which is
scaled to fit the figure.

\vskip 5mm

\noindent {\bf Figure 6.}
Comparison of the first derivative of different theoretical $m(M_{\rm
bol})$ relations: D'Antona \& Mazzitelli (1994), dotted curve; Baraffe
et al. (1995), long-dashed curve; our new models, solid line; and old
models, short-dashed line.  The low-resolution histogram is the
photometric luminosity function estimated by Kroupa (1995b) and scaled
to fit the figure. The Kroupa, Tout \& Gilmore (1993) empirical
$m(M_{\rm bol})$ relation (Table~A) is plotted as a histogram with
higher resolution.

\vskip 5mm

\noindent {\bf Figure 7.}
The Hertzsprung-Russell diagram for low-mass stars.  The observational
data compiled by Popper (1980) are represented by open symbols
corresponding to the masses given in the legend.  The empirical
estimates of Leggett et al. (1996) are in good agreement with the
Popper (1980) data.  The data published by Kirkpatrick et al. (1993)
are shown by filled circles. Various lines illustrate the stellar
models. Transverse dotted lines separate their mass ranges (from
bottom to top: 0.1, 0.2, 0.4, 0.6 and $1.0\,m_\odot$). 

\vskip 5mm

\noindent {\bf Figure 8.}
The first derivative of the theoretical $m(M_{\rm bol})$ relation for
metal abundances $Z=0.02$ (solid line), $Z=0.01$ (long-dashed line),
$Z=0.001$ (dash--dotted line), $Z=0.0003$ (short-dashed line) and
$Z=0.0001$ (dotted line).

\vskip 5mm

\noindent {\bf Figure 9.}
The relative position of the minimum in the first derivative of the
$m(M_{\rm I})$ relation as a function of metallicity.  The solid
circles are the bolometric magnitudes of our models shown in Fig.~8.  The
open squares connected by dotted lines bracket the position of the
maximum in the theoretical $-dm/dM_{\rm I}(M_{\rm I})$ relations
computed by D'Antona \& Mazzitelli (1996) after shifting the maxima by
$+0.7\,$mag to account for the offset of the peak from the empirical
position (see Fig.6).  Each open circle represents the absolute I-band
magnitude of the peak or turn-over in the observed luminosity
functions of different stellar clusters. Errorbars indicate the
bin-width in magnitudes used to construct the luminosity
function. Thick symbols represent stellar luminosity functions based
on more than a few hundred stars and thin symbols show luminosity
functions that are less well defined. The solar-neighbourhood
photometric luminosity function is labeled by ``SN''. The position of
the Pleiades luminosity function is shown from top to bottom after
adjustment for the following combinations of (distance modulus, age):
(5.5, $7\times10^7$~yr), (5.5, $1.2\times10^8$~yr), (5.5, main
sequence), (6, $7\times10^7$~yr), (6, $1.2\times10^8$~yr), (6, main
sequence).  

\vfill \bye